# Comparative study of panel and panelless-based reflectance conversion techniques for agricultural remote sensing

**Baabak G. Mamaghani[a], Carl Salvaggio[a]**
[a]Rochester Institute of Technology, College of Science, Chester F. Carlson Center for Imaging Science, 54 Lomb Memorial Drive, Rochester, NY 14623, USA

**Abstract.** Small unmanned aircraft systems (sUAS) have allowed for thousands of aerial images to be captured at a moments notice. The simplicity and relative low cost of flying a sUAS has provided remote sensing practitioners, both commercial and academic, with a viable alternative to traditional remote sensing platforms (airplanes and satellites). This paper is an expanded follow-up study to an initial work. Three radiance-to-reflectance conversion methods were tested to determine the optimal technique to use for converting raw digital count imagery to reflectance maps. The first two methods utilized in-scene reflectance conversion panels along with the empirical line method (ELM), while the final method used an upward looking sensor that recorded the band-effective spectral downwelling irradiance. The methods employed were 2-Point ELM, 1-point ELM, and At-altitude Radiance Ratio (AARR). The average signed reflectance factor errors produced by these methods on real sUAS imagery were: -0.005, -0.0028, and -0.0244 respectively. These errors were produced across four altitudes (150, 225, 300 and 375ft), six targets (grass, asphalt, concrete, blue felt, green felt and red felt), five spectral bands (blue, green, red, red edge and near infrared), and three weather conditions (cloudy, partly cloudy and sunny). Finally, data was simulated using the MODTRAN code to generate downwelling irradiance and sensor reaching radiance to compute the theoretical results of the AARR technique. A multitude of variables were varied for these simulations (atmosphere, time, day, target, sensor height, and visibility), which resulted in an overall theoretically achievable signed reflectance factor error of 0.0023.

**Keywords:** small unmanned aircraft systems, sUAS, calibration, reflectance panels, radiance, reflectance, MODTRAN, MicaSense RedEdge-3, reflectance conversion, agricultural science, agriculture

1. Introduction
    1.1. Small Unmanned Aircraft Systems (sUAS)

   Development of sensors and cameras for observing the Earth have always been at the forefront of advancements in technology in the remote sensing community. Previous platforms were constrained to satellites and manned aircraft, which made collections expensive and resulted in data with relatively large ground sampling distances (GSDs). This limited the applications of remote sensing to particular scales and collection/revisit frequencies. Recently, improvements have been made to the platforms that carry the sensors and cameras [1]. The enhancements in these platforms has led to affordable sUAS for both civil and research applications [2] [3], which have led to a large influx of data with high spatial and temporal resolution [4]. This has been a big assistance for in agricultural remote sensing, because an increase of higher quality data has been able to provide better soil, crop, and pest maintenance

[5]. With the ability to collect data at any moment in time, it is important that everybody using a sUAS for scientific purposes knows the correct methods to process the raw imagery that is captured to an invariant ground reflectance factor.

### 1.2. Reflectance Conversion

Various factors affect image data that are captured remotely. The atmosphere, time of day, location, and sensor altitude are some of those variables. Before two remotely sensed images can be compared to one another, reflectance conversion, which is defined as conversion of radiance into reflectance using a measured incident radiance and ground leaving radiance, needs to occur to ensure that the images are being represented in the same domain. To show the importance of reflectance conversion, four mosaics were generated using Agisoft PhotoScan [6] and displayed in Figure 1. The top two mosaics are radiance and reflectance generated without mosaic blending. Without converting the images to reflectance, the radiance images show prominent variation in illumination that was incident during the approximate 20-minute duration of the sUAS flight. After conversion to reflectance, the mosaic demonstrates a smoother appearance which is a result of conversion to an illumination invariant space, reflectance factor. The bidirectional reflectance distribution function (BRDF) was mitigated by using images with reflectance conversion panels within 10° of nadir.

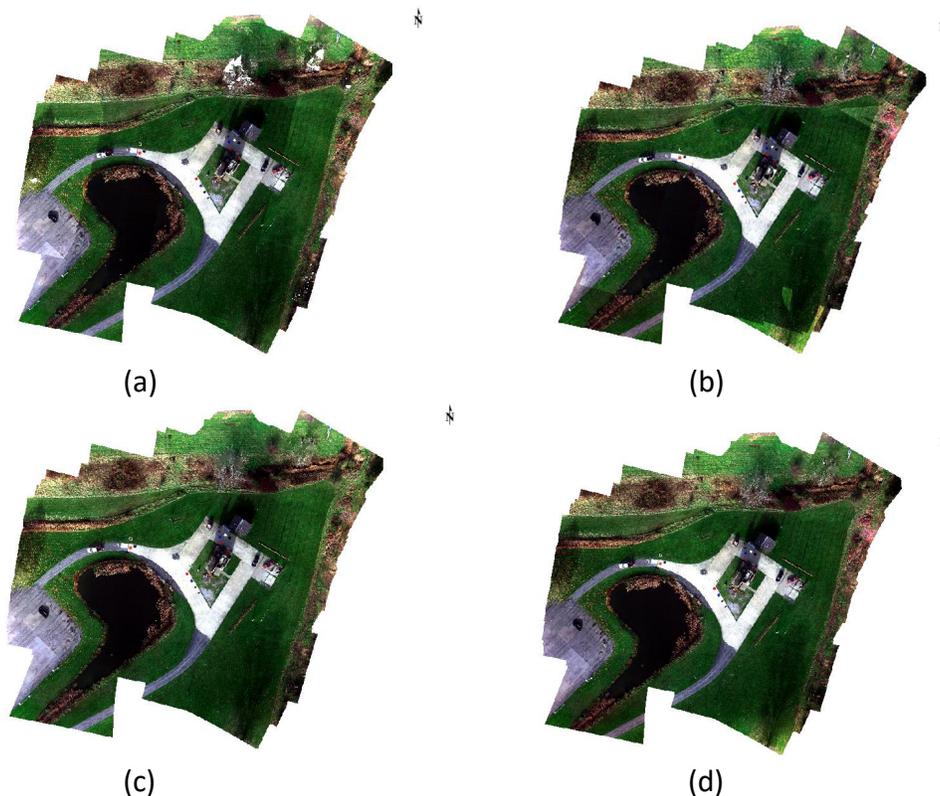

**Figure 1:** Image mosaics derived under partly cloudy conditions without blending using (a) radiance imagery, (b) reflectance imagery and with mosaic blending using (c) radiance imagery and (d) reflectance imagery. The imagery was collected using a MicaSense RedEdge-3 sensor for an altitude of 150ft on November 9, 2017.

This ensured that the mosaic had been converted to an illumination invariant space. The bottom two mosaics were generated using radiance and reflectance imagery along with mosaic blending, which ``implements an approach with data division into several frequency domains which are blended independently. The highest frequency component is blended along the seam line only, each further step away from the seam line resulting in a less number of domains being subject to blending" [7]. This method is not recommended.  Even though the mosaic appears visually more appealing, the seam line blending creates pixel values that were never physically recorded, which can result in incorrect analysis. Figure 2 displays the non-blended mosaics as Normalized Difference Vegetation Index (NDVI) maps created from radiance and reflectance factor, a metric used to determine the health of vegetation [8] [9], to demonstrate the impact of reflectance conversion for a common application of remote sensing data in precision agriculture.

It is very important for all remote sensing practitioners to know how prepackaged software handles data. Without mosaic blending, the data was displayed as loaded into Agisoft PhotoScan. When mosaic blending is turned on, the images are altered to be visually similar with their neighbor images, which is an incorrect way of overlaying various remotely sensed images. Images should be converted to reflectance before they are compared against one another, and should be converted to reflectance using the proper methods. Figure 3 shows the mosaic blending issue. Software packages, like PhotoScan, will take all images that overlap and create some combination between all the pixels. If the images were in reflectance space, they would all, theoretically, have the same value and no combination would be necessary.

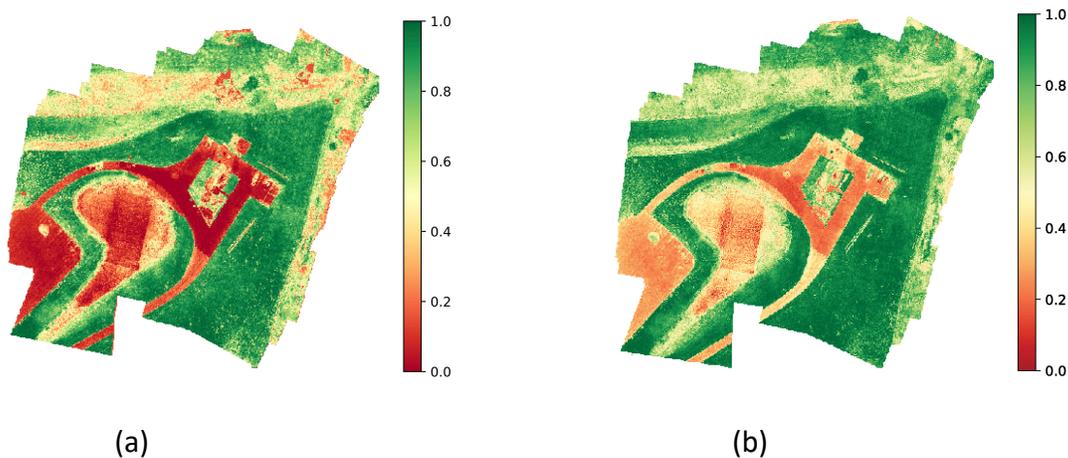

(a)          (b)

**Figure 2:** NDVI maps derived with image mosaics derives under partly cloudy conditions without blending using (a) radiance imagery and (b) reflectance imagery. The imagery was collected using a MicaSense RedEdge-3 sensor for an altitude of 150ft on November 9[th], 2017.

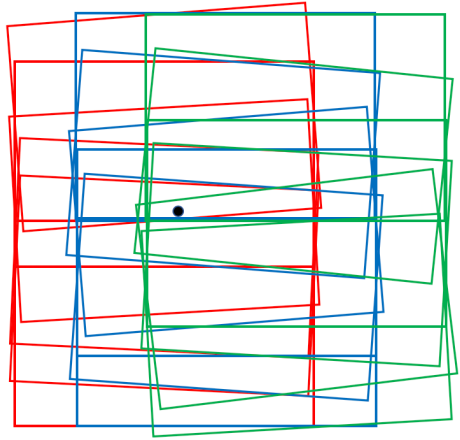

**Figure 3:** Example overlay of sUAS imagery captured from a flight (shot polygon). Colors represent different flight lines. The black dot represents a single point on the ground which was imaged by all polygons. This point will have different digital counts and radiance values in each of the images, but will have the same reflectance value.

### 1.3. Motivation

The motivation for this study developed during numerous sUAS data collections during the 2017 calendar year. Every sUAS collect required reflectance conversion targets to be placed in the field, and the targets needed to be revisited by the sUAS multiple times during the individual flights. With these task requirements, less area was able to be covered by the sUAS during any flight, and the physical process consumed valuable time in the field for the ground crew. In addition, the illumination of the field/targets could constantly be changing, especially in partly cloudy conditions. Therefore, ground reference data had to be collected and time stamped as frequently as possible, in order produce the most accurate reflectance imagery. This method (and others like it) are typically referred to as ``Scene-Based Empirical Approaches'' [10].

Another widely used method of converting radiance imagery into reflectance is: radiative transfer modeling. Radiative transfer models such as Fast Line-of-sight Atmospheric Analysis of Spectral Hypercubes (FLAASH), Atmospheric Removal (ATREM), Atmosphere Correction Now (ACORN), and High-accuracy Atmospheric Correction for Hyperspectral Data (HATCH) have been developed and used. These methods compute the atmospheric effects based on the physical radiative transfer model and the parameters of the atmosphere at the time the image/data was recorded [11] [12] [13]. Some atmospheric conditions (such as water vapor) are difficult to measure at the time of flight and are therefore estimated. In addition, some parameters are spatially varying across the areal extent of the study area [14]. Finally, these methods have mainly been utilized for images captured at traditional remote sensing altitudes (via satellites and airplanes). However, there is very little atmospheric influence between the ground and the sUAS, which allows users to use other techniques for reflectance conversion.

Due to the minimal atmospheric effects over the path ranges seen from allowable sUAS collection altitudes, a simple method for converting images from radiance into reflectance for sUAS is possible. This would save significant time in the field, as well as post-processing time. This method would be helpful to not only researchers, but any sUAS remote sensing practitioner (i.e. farmers, land surveyors, and insurance companies). The work in this paper looks at an in-depth study into a technique that can revolutionize sUAS remote sensing.

### 1.4. Agricultural Impact

Improving accuracy of data has long been addressed by remote sensing scientists to allow for proper decisions to be made against important problems such as crop yield, oil exploration, and global climate change. By calibrating all captured images to obtain reflectance, the data becomes independent of time, illumination, and weather conditions. The effect of calibration for imagery collected under different illumination conditions can be seen in Figures 1 and 2. By correcting the differently illuminated digital count values into radiance and then into reflectance, a better representation of the targets is produced. This is noticeable when the images are stitched together. If images are captured on a partly cloudy day, the illumination difference between images is noticeable, but after conversion to reflectance, the images seamlessly stitch together. If the same crop is imaged on multiple sUAS flight paths, (as seen in Figure 3), the illumination variations between these images can result in various vegetation health indices being computed. By calibrating for illumination, weather, and time, the standard deviation of the measured health index for that particular crop will be significantly reduced, and they will be a more accurate representation of that crop. This is of vital importance for agricultural scientists and anyone else who deals with agricultural products. Using sUAS remote sensing technologies and techniques, the time needed to analyze a crop field can drastically be diminished compared to ground based mechanical methods of physically walking the field, but this increases the importance for calibration if this new technology is to fit in to a farmer's already busy schedule. If a farmer's crop field is imaged and analyzed without any calibration, true negative, and false positive results can occur. In the true negative case, the analysis would show the crops are unhealthy and need to be attended to. This causes the farmer to spend time and resources to ensure their crops are healthy. In the false positive case, the analysis would show a healthy crop field and the farmer would not worry when in fact, the crops are in some way unhealthy. This will lead to further damage to the crops and potentially a loss of a yield and potential profit. These two scenarios, and many more, are why calibration is very important for agricultural science and research.

## 2. Background

### 2.1. Empirical Line Method

The Empirical Line Method (ELM) is a technique used to convert remotely sensed images from at-sensor digital count (or radiance) to at-surface reflectance [15]. By having a large object of known reflectance in-scene, an image can be converted to reflectance using the linear relationship shown by Equation 1,

$$\rho_i = m_i L_{s,i} + b_i \quad (1)$$

where, $\rho_i$ is the reflectance factor, $m_i$ is the slope, $L_{s,i}$ is the band effective spectral radiance, $b_i$ is the bias, with $i$ denoting the spectral band number. The slope and bias are derived by using two reflectance conversion panels in the scene, and measured ground reference reflectance spectra to map radiance to reflectance. For optimal results, it has been recommended to use two reflectance conversion targets (one bright and one dark) [16] [17] [18] [19] [20], make reflectance measurements as close to data collection time as possible, and ensure the reflectance conversion targets are level (perpendicular to sensor nadir) [21]. Others have used four reflectance conversion targets to calculate the slope and offset to improve the accuracy of the linear relationship [22] [23]. It is important to note that while the linear relationship between radiance and reflectance holds, the slope and bias values will change for each waveband. Therefore, ELM has to be accomplished separately for every spectral band produced for a given sensor. ELM has been widely used in the field of remote sensing for reflectance conversion. It has been used to retrieve reflectance factors from satellites such as Landsat [24] and IKONOS [25], has been utilized for mineral mapping [18], and even vegetation classification [20].

### 2.2. At-altitude Radiance Ratio (AARR)

A new method for converting radiance to reflectance was discussed by Mamaghani et al. in 2018 [26]. By dividing the radiance imagery captured from the sUAS by the downwelling radiance onto the sUAS, reflectance factor imagery can be produced (with an average absolute reflectance factor error of 0.0287). Equation 2 demonstrates this relationship.

$$\rho_i = \frac{L_{s,i}}{\frac{E'_{solar,i}}{\pi} \cos(\sigma') \tau'_i + L_{\downarrow,solar,i}} \qquad (2)$$

where, $\rho_i$ is the reflectance factor, $L_{s,i}$ is the band effective spectral radiance, $E'_{solar,i}$ is the spectral exoatmospheric solar irradiance, $\sigma'$ is the solar zenith angle, $\tau'_i$ is the spectral transmission from space to the sUAS, $L_{\downarrow,solar,i}$ is the solar scattered downwelling sky radiance propagating on to the sUAS, and $i$ denotes the spectral band number.

Others have tried similar techniques. Hakala et al. applied this technique for their hyperspectral images captured from an sUAS with the use of a downwelling irradiance spectrometer [27]. They concluded that this technique was ``extremely attractive, as it simplifies the field operations, and it is suitable for operations in varying illumination conditions...". In addition, they believed that further studies can evaluate the quality of this reflectance conversion technique by using various targets and brightness levels which was the motivation for this paper.

In addition, Lekki et al. computed spectral reflectance by ratioing the radiance leaving water and the downwelling irradiance which showed ``...good qualitative agreement between reflectance measured at the surface and the airborne measurement" [28]. Ortiz et al. tested three different ELM techniques, along with a fourth technique where they divided the at-sensor radiance with the downwelling irradiance collected by attaching a spectroradiometer (ASD FieldSpec HH2 with cosine theta receptor) on top of the aircraft [29]. Their results showed that the ELM techniques utilized had the best estimate of absolute reflectance.

### 2.3. Previous Initial Study

An initial study was accomplished by Mamaghani et al. on this topic [26]. Their results showed 2-Point ELM produced the smallest mean absolute error in band effective reflectance factor (0.0165), and that AARR was a reliable technique in producing reflectance images (0.0287). Despite these findings, they analyzed the errors of only three targets (asphalt, grass, and concrete) from a very small sample size (five images from a single sUAS flight). In addition, the simulations they conducted to validate AARR only permuted the height of the sensors (2, 150, 225, 300, 375 and 5000ft), and the targets they observed (asphalt, grass, and concrete).

The work presented in this paper expands that previous study. Over 1000 images were analyzed across 12 sUAS flights; four flight altitudes (150, 225, 300 and 375ft) under three different sky conditions (cloudy, partly cloudy and sunny). Each method tested in this paper analyzed 1,820 targets across those images (images may have had multiple targets). Finally, simulations were run to model how AARR performs under a new variety of image acquisition conditions including atmosphere type, time, season and visibility.

### 2.4. MicaSense RedEdge-3
#### 2.4.1. Camera and Downwelling Light Sensor

MicaSense has assisted agricultural experts to improve their crop management by developing the RedEdge-3 camera, which is a multispectral sensor that captures five bands simultaneously. In addition, MicaSense offers a downwelling light sensor (DLS) that is designed to capture the downwelling irradiance per band per image. This DLS sensor measures light incident on a diffuser, providing a nearly complete view of the hemisphere above the detector surface [30]. An irradiance fall-off test was carried out to confirm MicaSense's DLS hemispherical FOV claim as seen in Figure 10. Both the camera and the DLS can be mounted onto a sUAS, which allows the user to capture both the image and the illumination conditions simultaneously. One of the reflectance conversion techniques presented in this paper utilizes the DLS to convert radiance imagery to reflectance directly. Table 1 presents the five spectral bands with their wavelengths and bandwidths [30].

**Table 1:** MicaSense RedEdge-3 spectral bands with respective center wavelengths and bandwidth values.

| Band Name | Wavelengths [nm] | FWHM [nm] |
|---|---|---|
| Blue | 475 | 20 |
| Green | 560 | 20 |
| Red | 668 | 10 |
| Red Edge | 717 | 10 |
| NIR | 840 | 40 |

Every captured image contains metadata values to assist in converting digital count to radiance (W/m^2/sr/nm). The metadata values used in this approach include a vignette correction function, radiometric calibration coefficients, sensor gain, exposure time, and black level values.

One of the more recent studies into radiometric calibration of sUAS multispectral cameras tested a MicaSense RedEdge sensor both in lab and in the field. Their study concluded that while there is a high linear correlation in the lab between radiance measurements of the RedEdge and an ASD (Analytic Spectral Device - a calibrated hyperspectral scanner), the field measurements suggested that sUAS users should be cautious with the raw reflectance imagery that is produced [31].

### 2.4.2. Relative Spectral Response

To accurately model the output of the MicaSense RedEdge-3 used in this study, the relative spectral response (RSR) functions were measured. To calculate these curves, two spectral measurements were taken with a Newport Model 74004-1 monochromator.

The first spectral measurement was the spectral radiant flux, or spectral power, $\Phi(\lambda)$ [W]. A power meter was placed directly in front of the exit port of the monochromator, and measurements were made from 400nm to 900nm in 2nm increments. Each wavelength increment had a delay to allow for stabilization of the monochromator, which reduced noise in the values being recorded.

The second spectral measurement utilized the MicaSense RedEdge-3 camera to capture images of the radiance energy leaving the exit port of the monochromator. Images were manually captured over the same spectral range as the power measurements, at 2nm increments.

For each of the spectral bands of the camera, $i$, the peak normalized relative spectral response functions were computed as:

$$DC_{norm,i}(\lambda) = \frac{\overline{DC_i(\lambda)}}{g_i t_i} \qquad (3)$$

$$RSR_i(\lambda) = \frac{0.9975(DC_{norm,i}(\lambda) - b)}{\Phi(\lambda)} \qquad (4)$$

$$RSR'_i(\lambda) = \frac{RSR_i(\lambda)}{max(RSR_i(\lambda))} \qquad (5)$$

where $DC_{norm,i}(\lambda)$ is the average digital count over a region of interest (ROI), $g_i$ is the gain, $t_i$ is the exposure time, $b$ is a shift factor that is equal to the lowest non zero value of $RSR_i(\lambda)$, and $\Phi(\lambda)$ [W] is the spectral radiant flux. The peak normalized relative spectral response functions for the MicaSense RedEdge-3 sensor used in this study (SN:1713165) can be seen below in Figure 4.

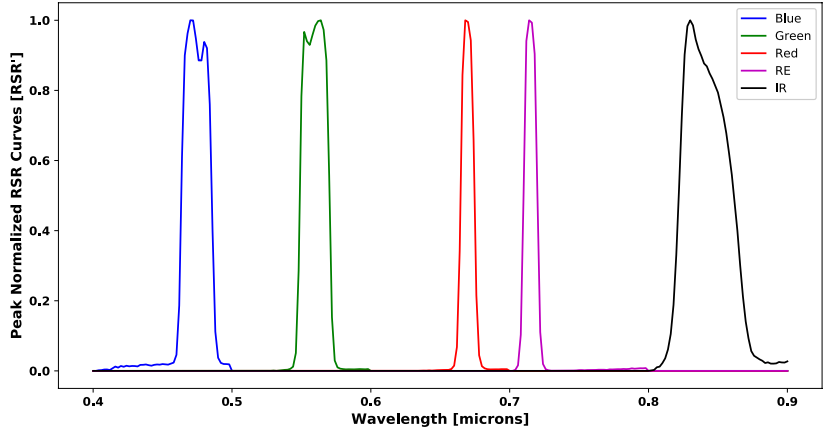

**Figure 4:** Peak normalized relative spectral response of the MicaSense RedEdge-3 used in study.

### 2.4.3. Sensor Reaching Radiance

As described in [32], the spectral radiance reaching a sensor can be written as:

$$L_s = \left[\frac{E'_{solar}(\lambda)}{\pi}\cos(\sigma')\tau_1(\lambda)\rho(\lambda) + L_{\downarrow,solar}(\lambda)\rho_d(\lambda)\right]\tau_2 + L_{\uparrow,solar}(\lambda) + L_a(\lambda) \quad (6)$$

where, $E'_{solar}(\lambda)$ is the spectral exoatmospheric solar irradiance, $\sigma'$ is the solar zenith angle, $\tau_1(\lambda)$ is the spectral transmission from space to the target, $\rho(\lambda)$ is the spectral directional reflectance function for the target, $L_{\downarrow,solar}(\lambda)$ is the solar scattered downwelling sky radiance propagating on to the target, $\rho_d(\lambda)$ is the spectral diffuse reflectance for the target, $\tau_2$ is the spectral transmission from target to sensor, $L_{\uparrow,solar}(\lambda)$ is the solar scattered spectral path radiance generated in the path between target and the sensor, and $L_a(\lambda)$ is the spectral scattered radiance from scattered photons due to background objects near the target of interest (adjacency). Equation 2 does not include the variables $\tau_2$, $L_{\uparrow,solar}(\lambda)$, or $L_a(\lambda)$ in the denominator because the short path (150-375ft) has negligible transmission loss, solar scattered spectral path and spectral scattered radiance.

In order to compute band effective radiance, $L_{s,i}$, the spectral radiance that reaches the sensor needs to be integrated over the appropriate spectral bandpass that is defined by the relative spectral response functions. Therefore,

$$L_{s,i} = \frac{\int_\lambda L_s(\lambda)RSR'_i(\lambda)d\lambda}{\int_\lambda RSR'_i(\lambda)\,d\lambda} \quad (7)$$

where, $RSR'_i(\lambda)$ is the peak normalized relative spectral response for the $i^{th}$ band.

## 3. Methodology

Three different reflectance conversion methods were compared - 2-point empirical line method (2-point ELM), 1-point empirical line method (1-point ELM) and at-altitude radiance ratio (AARR). To convert the raw sUAS images into reflectance images, a two step process was required. The first step converted the imagery from raw digital count to radiance and the second step converted radiance to reflectance.

### 3.1. Digital Count to Radiance

In order to make comparisons between the reflectance conversion methods, it was first necessary to convert the images from raw digital count to radiance. This employed method was developed and made available in open source software by MicaSense [33]. The equations below depict the process of converting raw images into radiance.

$$L_{s,i} = \frac{I_{cor,i}}{g_i t_i} \frac{a_{1,i}}{2^N} \tag{8}$$

where, $L_{s,i}$ is the band effective radiance image, $I_{cor,i}$ is the corrected image, $g_i$ is the gain (1, 2, 4 or 8), $t_i$ is the exposure time [μs], $a_{1,i}$ is a radiometric calibration coefficient, $N$ is the bits per pixel, and $i$ denotes the spectral band number. Equation 9 represents the corrected image $I_{cor,i}$,

$$I_{cor,i} = V_i R_i (I_{raw,i} - dL) \tag{9}$$

where, $V_i$ is the vignette correction function map (Equation 10) $R_i$ is the CMOS array readout correction image (Equation 13), $I_{raw,i}$ is the raw digital count image, $dL$ is the dark level, and $i$ denotes the spectral band number. The vignette map, $V_i$, is represented as

$$V_i = \frac{1}{k} \tag{10}$$

where $k$ is a correction factor

$$k = 1 + k_0 r + k_1 r^2 + k_2 r^3 + k_3 r^4 + k_4 r^5 + k_5 r^6 \tag{11}$$

where $k_0$ through $k_5$ are polynomial correction coefficients, and $r$ is the distance of the pixel to the vignette center

$$r = \sqrt{(x - c_x)^2 + (y - c_y)^2} \tag{12}$$

where $c_x$ and $c_y$ represent the vignette center. The CMOS array readout correction image, $R_i$, is

$$R_i = \frac{1}{1 + a_{2,i} \frac{y_i}{t_i} + a_3 y_i} \tag{13}$$

where $a_{2,i}$ and $a_{3,i}$ are radiometric calibration coefficients, $y_i$ is the pixel row number, and $i$ denotes the spectral band number.

### 3.2. Radiance to Reflectance

Before converting all the images acquired during a sUAS flight from radiance to reflectance, a subset of the images from that flight were tagged as reflectance conversion images. Reflectance conversion images were defined as images with both a bright and dark reflectance conversion panel that fell within 10° of nadir. When an image was being converted to reflectance, a Euclidean distance was computed between the current image's DLS irradiance vector (a five element vector, one irradiance value per band) and the DLS irradiance array of every identified reflectance conversion image. The reflectance conversion image which is closest in spectral downwelling radiance space, that is most similar in illumination, was selected for use in converting that particular image.

The DLS radiance, as shown in the denominator of Equation 2, is related to the DLS irradiance by a factor of $\pi$, and shown in Equation 14.

$$E_{\downarrow,solar} = \pi \left[ \frac{E'_{solar}}{\pi} \cos(\sigma') \tau' + L_{\downarrow,solar} \right] \tag{14}$$

where, $E_{\downarrow,solar}$ is the recorded irradiance value from the MicaSense RedEdge-3's DLS, $E'_{solar}$ is the spectral exoatmospheric solar irradiance, $\sigma'$ is the solar zenith angle, $\tau'$ is the spectral transmission from space to the sUAS, $L_{\downarrow,solar}$ is the solar scattered downwelling sky radiance propagating on to the sUAS, and $i$ denotes the spectral band number. Equation 15 is used to determine which reflectance conversion image should be used.

$$d(E_{\downarrow,solar,img}, E_{\downarrow,solar,cal_k}) = \sqrt{\sum_{i=1}^{5} (E_{\downarrow,solar,img} - E_{\downarrow,solar,cal_k})^2} \tag{15}$$

where, $d(E_{\downarrow,solar,img}, E_{\downarrow,solar,cal_k})$ is the Euclidean distance between the DLS irradiance vector of the image being converted to reflectance, $E_{\downarrow,solar,img}$, and the DLS irradiance vector of the current reflectance conversion image, $E_{\downarrow,solar,cal_k}$. $K$ is the number of reflectance conversion images for the current sUAS flight, and $i$ denotes the spectral band of DLS measurement.

This new method of selecting the reflectance conversion image is preferable to a single reflectance conversion image collected at the beginning or the end of the flight, because every image is converted to reflectance using a linear relationship that is developed from reflectance conversion panels that were imaged under the most similar illumination conditions to the current image.

In addition to using this new method of reflectance conversion image selection (DLS irradiance vector comparison), two other methods will be evaluated with 1-point and 2-point ELM: single image and time. Single image reflectance conversion utilizes a single image of the reflectance conversion targets to convert every image captured during the collection, while time-based conversion uses the reflectance conversion targets closest in time to the current image for reflectance conversion. These two methods will be compared with the newly developed DLS comparison method.

### 3.2.1. 1-Point ELM

A linear relationship between radiance and reflectance can be formed by using 1-point ELM. A single point (bright panel) and the origin are used to determine the conversion factor (slope). The reflectance conversion image to use was selected using the Euclidean distance formula in Equation 15.

$$\rho_i = m_i L_{s,i} \tag{16}$$

where, $\rho_i$ is the reflectance factor, $m_i$ is the slope, $L_{s,i}$ is the spectral radiance, and $i$ denotes the spectral band number. Equation 17 represents the slope, $m_i$,

$$m_i = \frac{p_{i,bright}}{L_{s,i,bright}} \tag{17}$$

where $p_{i,bright}$ is the average ground reference reflectance of the bright reflectance conversion panel, and $L_{s,i,bright}$ is the average radiance of the bright reflectance conversion panel, and $i$ denotes the spectral band number.

This implementation of 1-point ELM is assuming that a radiance of zero would mean a reflectance of zero ($L_{s,i}$ and $\rho_i = 0$). Since all drone images are analyzed from altitude, the path radiance ($L_a(\lambda)$) will apply a small bias to the linear relationship. The error from the path radiance bias can be mitigated by adding a second reflectance conversion panel to the ELM.

### 3.2.2. 2-point ELM

By replacing the origin point in 1-point ELM with a second reflectance conversion point (dark panel), a more accurate reflectance conversion can be achieved. This is true because an actual ground reference reflectance will be mapped to a radiance value, which will introduce an updated slope and a bias term, which can be thought of as representative the path radiance ($L_a(\lambda)$). The reflectance conversion image is again selected using Equation 15 to find the most similarly illuminated set of reflectance conversion panels. For 2-point ELM, the following equations are used:

$$\rho_i = m_i L_{s,i} + b_i \qquad (18)$$

where, $\rho_i$ is the reflectance factor, $m_i$ is the slope, $L_{s,i}$ is the spectral radiance, $b_i$ is the bias and $i$ denotes the spectral band number. The slope, $m_i$, is represented as

$$m_i = \frac{p_{i,bright} - p_{i,dark}}{L_{s,i,bright} - L_{s,i,dark}} \qquad (19)$$

where $p_{i,bright}$ is the average ground reference reflectance of the bright reflectance conversion panel, $p_{i,dark}$ is the average ground reference reflectance of the dark reflectance conversion panel, $L_{s,i,bright}$ is the average radiance of the bright reflectance conversion panel, $L_{s,i,dark}$ is the average radiance of the dark reflectance conversion panel, and $i$ denotes the spectral band number. The bias, $b_i$, is given by

$$b_i = p_{i,bright} - m_i L_{s,i,bright} \qquad (20)$$

where $p_{i,bright}$ is the average ground reference reflectance of the bright reflectance conversion panel, $m_i$ is the slope (Equation 19), $L_{s,i,bright}$ is the average radiance of the bright reflectance conversion panel, and $i$ denotes the spectral band number. The bright panel was used in the above equation, however, the same bias would be calculated if the average ground reference reflectance of the dark panel and the average radiance of the dark panel was used.

### 3.2.3. AARR

Investigated by Mamaghani et al. [26], this method calibrates images using the downwelling radiance that was recorded with the image. The MicaSense RedEdge-3's DLS records the downwelling irradiance that is incident on the sensor, which is then divided by $\pi$ to produce the downwelling radiance as shown in Equation 14.

While the drone is flying, the DLS is angled in the direction of the flight path as well as by any stabilization to flight that the aircraft is performing. This affects the results recorded by the sensor, therefore, the DLS recorded irradiance needs to be corrected for sensor orientation. This correction has been recently implemented by MicaSense [33]. This correction is required so the recorded DLS irradiance values are converted to the irradiance values that would have been recorded, had the DLS sensor been looking straight up, at that moment. This ensures that each image's DLS spectral irradiance value is an accurate representation of the downwelling hemispherical irradiance. For example, if the sUAS is flying away from the sun, then the DLS would record a lower value than if the drone was flying towards the sun, at the same location in the sky. By correcting the DLS to look upward, both of those images would record the same values. Equation 21 depicts the corrected spectral downwelling irradiance

$$E_{\downarrow,solar,i} = \frac{E_{\downarrow,solar,raw,i}\left(\frac{E_d}{E_s} + \sin(\theta_{SolarElev})\right)}{Fres\left(\frac{E_d}{E_s} + \cos(\theta_{SunSen})\right)} \tag{21}$$

where $E_{\downarrow,solar,raw,i}$ is the raw DLS spectral irradiance (measured at the time of the image capture), $\frac{E_d}{E_s}$ is the diffuse coefficient, $\theta_{SolarElev}$ is the solar zenith angle, $Fres$ is the Fresnel correction factor, $\theta_{SunSen}$ is the sun sensor angle, and $i$ denotes the spectral band number. A diffuse coefficient of $0.1\overline{66}$ is used because around 85% of the light is direct, with the other 15% being diffuse according to [33].

### 3.3. Modeled At-altitude Radiance Ratio (M-AARR)
#### 3.3.1. MODTRAN

To examine the theoretical capability of the AARR technique under a wide range of conditions that might be found around the world, simulations were run using Spectral Science Incorporated's (SSI) moderate resolution atmospheric transmission (MODTRAN4) code. MODTRAN is used ``... for the prediction and analysis of optical measurements through the atmosphere" [34]. The code simulates the propagation of electromagnetic energy through/from the atmosphere over a range of 0.2μm to 100μm. Theoretically, any scene can be simulated, because MODTRAN allows users to vary the atmospheric constituency, time, day, location, sensor height, and sensor orientation, among many other variables including target and background reflectance spectra.

#### 3.3.2. Setup

A MODTRAN simulation is specified by a single input file, tape5, which stores all of the simulation input parameters. For simplification, a script was written to modify the tape5 file for every set of parameter permutation. The permuted parameters are listed in Table 2.

One of the most relevant features of MODTRAN, for the purpose of this study, was the spectral albedo file (spec_alb.dat). The vendor-provided spec_alb.dat file contains reflectance curves of various generic targets, such as forest, ocean, desert, grass and various constant reflectances. Since comparisons between modeled and real data were desired, the ground reference reflectance curves of the six targets in our experiment scene (asphalt, grass, concrete, blue felt, green felt, and red felt) were added to the spec_alb.dat file and used in the simulations. This added authenticity to the simulations. Example reflectance factor curves of the ground reference targets, measured on November 9, 2017, can be seen in Figures 5-8.

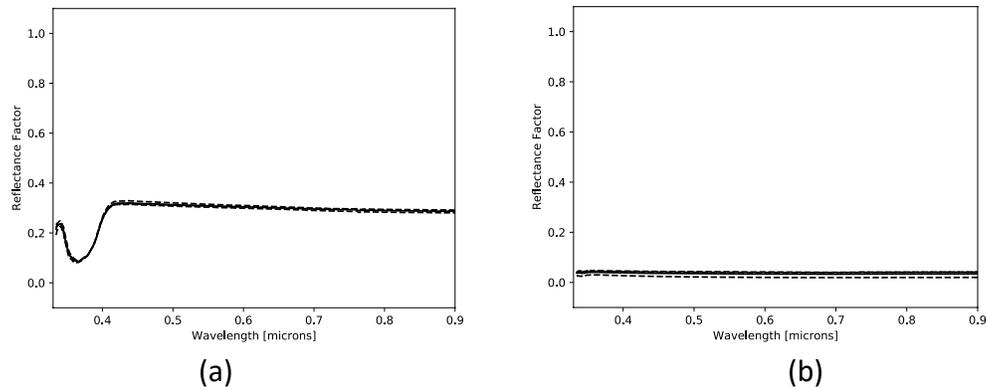

**Figure 5:** Collected (a) bright panel and (b) dark panel spectra during sUAS flight on November 9, 2017. Dashed lines are single measurements, while solid lines are the average measurement.

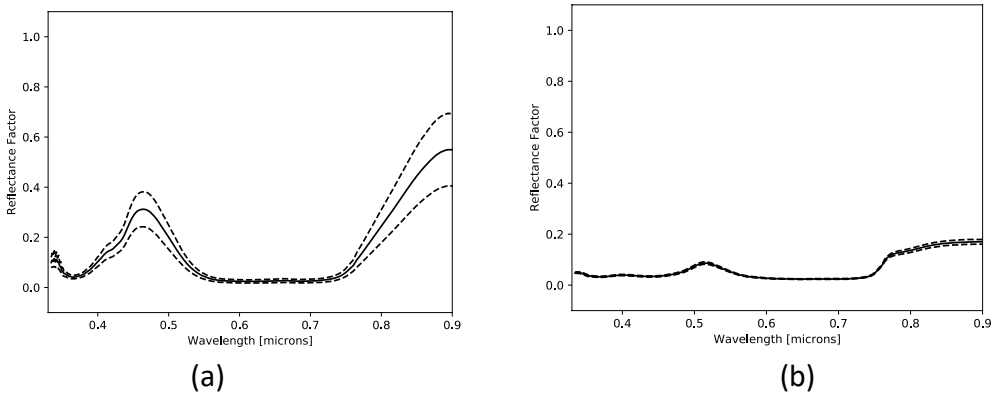

**Figure 6:** Collected (a) blue felt and (b) green felt spectra during sUAS flight on November 9, 2017. Dashed lines are single measurements, while solid lines are the average measurement.

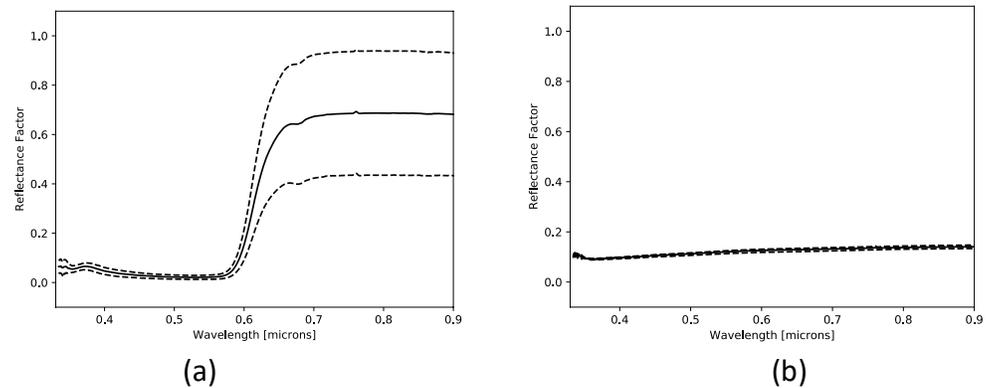

**Figure 7:** Collected (a) red felt and (b) asphalt spectra during sUAS flight on November 9, 2017. Dashed lines are single measurements, while solid lines are the average measurement.

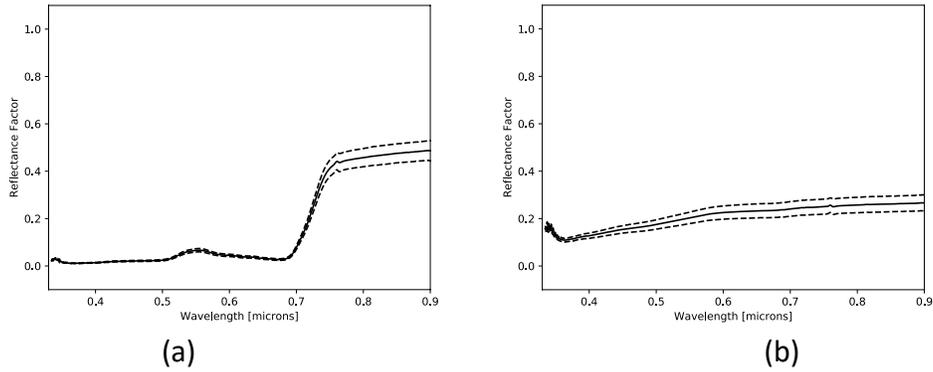

(a) (b)

**Figure 8:** Collected (a) grass and (b) concrete spectra during sUAS flight on November 9, 2017. Dashed lines are single measurements, while solid lines are the average measurement.

When the simulation concludes, MODTRAN creates an output file (tape7.scn). This file contains the computed spectral radiance, irradiance and transmittance measurements through the atmosphere. To model AARR, the components representing the total radiance observed by the sensor (TOTAL_RAD) and the ground reflected radiance (GRND_RFLT) are needed.

To produce at-altitude reflectance curves with these MODTRAN output parameters, two separate simulations were necessary. The first modeled the sensor reaching radiance, $L_s(\lambda)$, by aiming the sensor directly down at the target. The total radiance that is generated from this simulation was used as the sensor reaching radiance. The second modeled the DLS downwelling radiance. To correctly model this component, a 100% spectrally constant Lambertian reflector was placed at the same sensor altitude as used in the first simulation, and the sensor altitude was raised one meter above that. For this simulation, the ground reflected radiance, from the simulated Lambertian reflector, was used as the DLS radiance.

**Table 2:** Description and values of MODTRAN variables used in M-AARR simulations [35]

| Description | MODTRAN Variable | Value |
|---|---|---|
| **Model Atmosphere** | MODEL | 1 (tropical), 2 (mid latitude summer), 3 (mid latitude winter), 6 (US Standard) |
| **Path Type** | ITYPE | 2 (Slant or Vertical Path Between Two Altitudes) |
| **Surface Albedo** | SURREF | 'LAMBER' |
| **Surface Temperature [K]** | AATEMP | 303 |
| **Target** | CSLAB | Grass, Concrete, Asphalt, 100% Constant (from 'spec_alb.dat' |
| **Background** | CSLAB | Grass, Concrete, Asphalt (from 'spec_alb.dat') |
| **Visibility [km]** | VIS | 5.0, 10.0, 15.0, 23.0 |
| **Ground Altitude [km]** | GNDALT | 0.168 |

| Sensor Altitude [km] | H1 | 0.169, 0.214, 0.237, 0.259, 0.282, 1.692 |
|---|---|---|
| Target Altitude [km] | H2 | 0.168 |
| Day Number | IDAY | 79, 171, 265, 355 |
| Latitude | PARM1 | 43.041 |
| Longitude | PARM2 | 77.698 |
| Time [UTC] | TIME | 14.0, 15.0, 16.0, 17.0, 18.0 |
| Starting Wavelength [μm] | V1 | 0.33 |
| Ending Wavelength [μm] | V2 | 1.2 |
| Wavelength Increment [μm] | DV | 0.001 |
| FWHM [μm] | FWHM | 0.001 |

## 4. Data Collection

### 4.1. Location

Permission was granted to the RIT flight team to operate their sUAS system at the Henrietta Fire District Training Center Station Number 6 (43.041099°N 77.698343°W), near Rochester, NY, USA. The reflectance conversion targets used in the scene were two painted wooden panels (one dark and one bright), and a Labsphere-produced gray-scale portable fabric target [36] with three reflectance conversion bars of varying reflectance.

### 4.2. Variables

Three main variables were varied during data collection: sUAS height, weather, and test targets. The sUAS heights (150, 225, 300, and 375ft) were selected to demonstrate the effect of the atmosphere on reflectance conversion. 375ft was selected as the highest altitude because the Federal Aviation Administration of the United States of America limits sUAS pilots to 400ft above their target of interest [37]. In addition, three weather conditions were flown under: sunny, partly cloudy, and cloudy. Previously, it was required to collect data under sunny conditions, because satellites and aircrafts needed a clear line of sight to the targets. Because sUAS technology allows users to fly at any moment, it was desired to see the results of flying in conditions other than sunny. Depending on these results, sUAS users might not have to wait for clear days to collect data. Partly cloudy weather was selected to showcase the importance of reflectance conversion for remotely sensed images, and how some previously used methods might fail. Finally, the in-scene experimental targets used were three colored felts (red, green, blue), and demarcated areas of in-situ asphalt, concrete, and grass. These targets were selected because they provided a wide range of spectra morphologies.

### 4.3. sUAS

Data collection was accomplished using a DJI Matrice 100 quadcopter with a MicaSense RedEdge-3 attached (Figure 9).

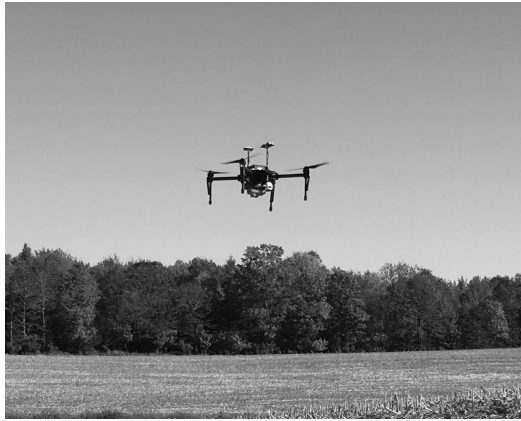
**Figure 9:** The DJI Matrice 100 quadcopter used for data collection

### 4.4. Ground Reference Measurements

Ground reference reflectance measurements were made to conduct reflectance conversion using both the 1-point and 2-point ELM as well as to use as ground reference data with which to compute errors in image-extracted reflectance factors for all experimental targets. A Spectra Vista Corporation (SVC) HR-1024i spectroradiometer was used to collect the reference data, which has a spectral range from 350-2500nm [38]. In order to collect accurate spectra, a reference spectrum was collected of a standard Spectralon™ panel before every target measurement to minimize any drift in scene-incident irradiance. Spectralon™ is a diffuse, spectrally flat material with a reflectance greater than 99% out to 1600nm [39] [40].

**Table 3:** MicaSense RSR band-integrated ground reference reflectance factors of in-scene targets recorded on November 8$^{th}$, 2017 (sunny conditions)

| Band [nm] | Grass | Asphalt | Concrete | Blue Felt | Green Felt | Red Felt |
|---|---|---|---|---|---|---|
| **Blue [475]** | 0.0219 | 0.1016 | 0.1794 | 0.3595 | 0.0541 | 0.2778 |
| **Green [560]** | 0.0693 | 0.1132 | 0.2318 | 0.0441 | 0.0535 | 0.0264 |
| **Red [668]** | 0.0264 | 0.1198 | 0.2602 | 0.0327 | 0.0298 | 0.6816 |
| **Red Edge [717]** | 0.1855 | 0.1217 | 0.2724 | 0.0375 | 0.0319 | 0.7205 |
| **NIR [840]** | 0.4912 | 0.1280 | 0.2957 | 0.4963 | 0.1808 | 0.7270 |

### 5. Results
#### 5.1. DLS FOV

As stated in Section 2.4.1, the DLS's FOV was tested with a simple setup using a light source and a collimator to validate MicaSense's hemispherical FOV claim. The collimator produced parallel light rays from the light source, which was the only source of illumination in the lab (all other sources of light were shut off). This allowed for the testing of the DLS falloff and the FOV by aiming the DLS directly at the light source, and then looking off axis. For every 10°, starting from 0° to 90°, a spectral irradiance measurement was captured with the DLS and then plotted in Figure 10 along with a cosine fall off. Cosine falloff was calculated by using the DLS spectral irradiances for 0° and multiplying by $\cos(\theta)$ for the remaining points. These results show that the DLS has a FOV close to the claimed 180°x180° and has the expected cosine fall-off factor.

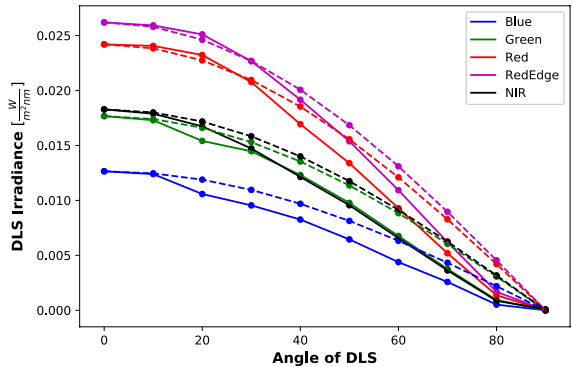

**Figure 10:** DLS spectral irradiance by angle. Collected spectra are the solid lines, while simulated cosine fall off are dashed lines.

### 5.2. sUAS Flight Results

Table 3 shows the band-integrated ground reference reflectance factors for all the in-scene targets on November 8, 2017. While it is highly unlikely that the reflectance of these experimental targets did not change between November 2, 2017 and November 9, 2017, other variables did. Data was captured on November 2, 8, and 9, 2017 because various weather conditions were desired for the flights. November 2$^{nd}$ was complete overcast, November 8$^{th}$ was sunny and November 9$^{th}$ was partly cloudy. In addition, variations in time of day (sun-target-sensor angle) and sky conditions (clear/partly cloudy/complete overcast) resulted in differences in band-integrated ground reference reflectance factors due to the bi-direction reflectance distribution factor for each of these targets. For this reason, it is critical that this ground reference data be collected as close as possible to the sUAS flight.

Overall results are shown in Table 4. They have been averaged over bands, targets, heights and atmospheres. 2-point ELM outperformed 1-point ELM and AARR for both signed (magnitude and direction) and absolute (magnitude only) errors. AARR standard deviation as also higher than both ELMs, which were very similar. For the ELMs, the dark wooden panel was used as the dark reflectance conversion point (~3%), and the middle bar of the gray-scale portable target was used as the bright reflectance conversion point (~30%).

**Table 4:** Overall average band effective reflectance factor errors with standard deviations. Both signed and absolute errors are shown.

| Error Type | AARR | 1-Point ELM | 2-Point ELM |
| --- | --- | --- | --- |
| Absolute | 0.0426 (0.050) | 0.0239 (0.0317) | 0.0219 (0.0335) |
| Signed | -0.0244 (0.0610) | -0.0028 (0.0396) | -0.0050 (0.0397) |

A few other results were investigated for the interest of the remote sensing community. The errors resulting from variations in sensor heights and atmosphere were examined more closely. The first analysis altered sensor altitude while averaging the error across bands, atmosphere, and targets. The second analysis altered atmosphere constituency while averaging the error across bands, sensor altitudes, and targets. For the purpose of the remaining studies, only the signed error results are presented.

Figure 11 shows the errors by sUAS altitude. All of these plots showed the same trend across conversion methods, a downward trend in average reflectance factor error from AARR to ELM. While all three methods produced similar results, ELM outperformed AARR with smaller average errors, and smaller standard deviations.

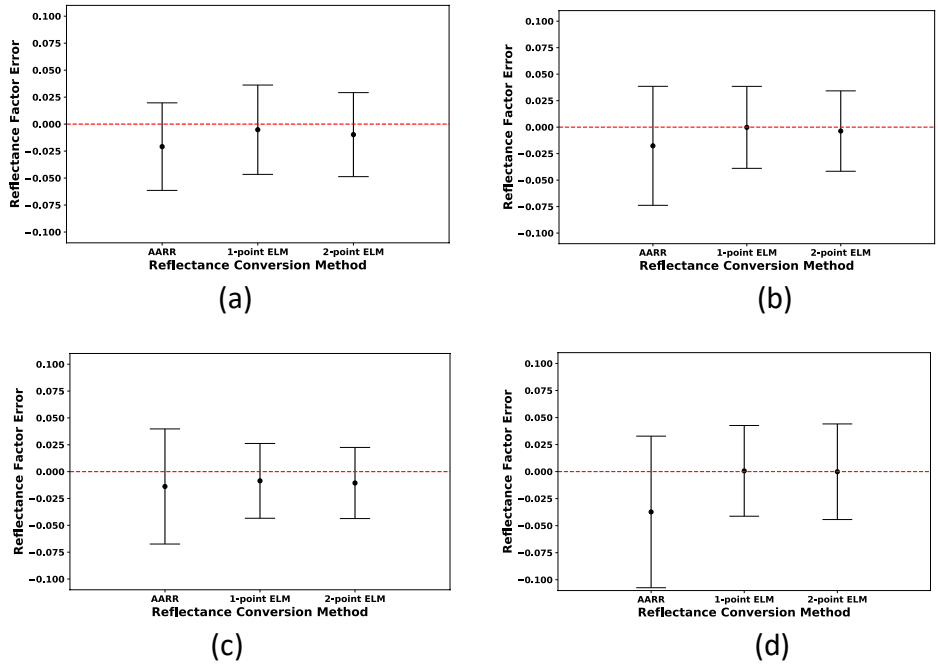

**Figure 11:** Signed reflectance factor errors averaged over MicaSense bands, atmosphere and targets for sensor altitude of (a) 150ft (b) 225ft (c) 300ft and (d) 375ft.

To further demonstrate how ELM outperformed AARR, Table 5 displays the signed band effective reflectance factor errors and standard deviations for each band of the MicaSense RedEdge-3. These errors have been averaged across all heights, atmospheres and targets. Across all bands, ELM had lower average errors and standard deviations.

**Table 5:** Signed band effective reflectance factor errors and standard deviations averaged across all heights, atmospheres, and targets.

|  | AARR |  | 1-Point ELM |  | 2-Point ELM |  |
|---|---|---|---|---|---|---|
|  | Mean | STD | Mean | STD | Mean | STD |
| **Blue** | -0.0290 | 0.0318 | 0.0019 | 0.0136 | -0.0044 | 0.0130 |
| **Green** | -0.0172 | 0.0341 | 0.0037 | 0.0164 | -0.0043 | 0.0152 |
| **Red** | -0.0291 | 0.0529 | 0.0016 | 0.0265 | -0.0065 | 0.0252 |
| **RedEdge** | -0.0175 | 0.0553 | -0.0059 | 0.0405 | -0.0130 | 0.0413 |
| **NIR** | -0.0294 | 0.1022 | -0.0151 | 0.0693 | 0.0030 | 0.0707 |
| **Overall** | -0.0244 | 0.0610 | -0.0028 | 0.0396 | -0.0050 | 0.0397 |

The atmosphere plots, in Figure 12, produced a more interesting result. The cloudy day (November 2, 2017) produced the smallest errors, even when compared to the sunny day. Since previous remote sensing platforms were satellites and manned aircraft, sunny (clear) conditions were required for image collection. While the signal-to-noise ratio for the recorded signal due to the radiance reaching the sensor is obviously higher on sunny days, the results in this study indicate that for a sUAS, cloudy weather produces smaller signed reflectance factor errors. Under sunny conditions, shadows can be present in the scene and energy can be reflecting off many surfaces back onto the targets which increases the potential for error without strenuous efforts to account for this scattered energy. Shadows are also very prominent in sUAS imagery due to the low flight altitude (high spatial resolution, small GSD). All days showed the expected trend in the reflectance factor errors were lower than AARR. Interestingly enough, the partly cloudy data produced reasonable errors when compared against cloudy and sunny. This is most likely due to the DLS being utilized to find the panels with the most similar illumination conditions.

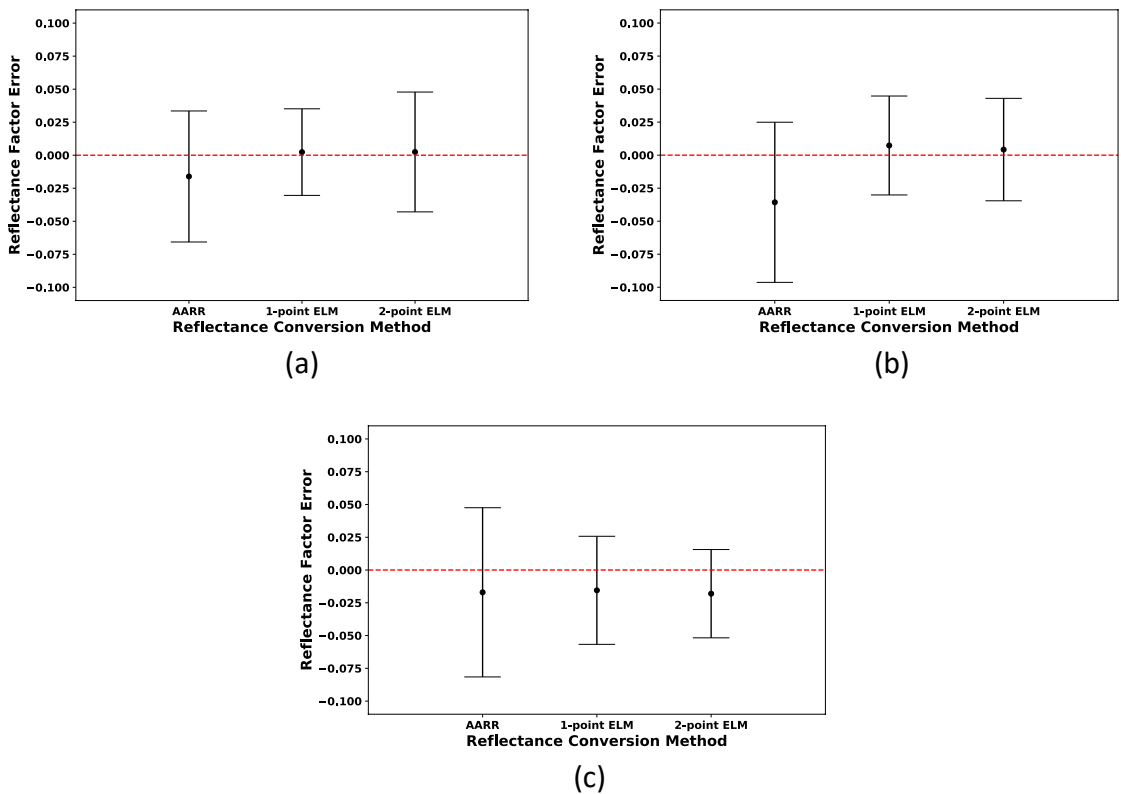

**Figure 12:** Signed reflectance factor errors averaged over MicaSense bands, sensor altitudes and targets for atmospheres of (a) cloudy (b) partly cloudy and (c) sunny.

### 5.3. DLS vs Single Image vs Time

Finally, the DLS method of reflectance conversion image selection was compared against single and time-based methods. Figure 13 displays the reflectance factor error box plots of all tested methods side by side. Of all ELM methods, DLS reflectance conversion image selection produced errors closest to zero. These results were expected because the Earth rotates and the sun changes position, which will either increase or decrease the irradiance onto the targets. As this occurs continuously throughout the sUAS flight, it is necessary to have more images captured of the reflectance conversion targets to most accurately represent the current illumination condition. Single image reflectance conversion does not take this issue into account, hence the higher error. Time based reflectance conversion is also flawed because the illumination of a scene can change in a few seconds. This would result in the wrong reflectance conversion panel image being used simply because it was captured a few seconds before. However, this did not occur during our data collections as can be seen by the time results in Figure 13.

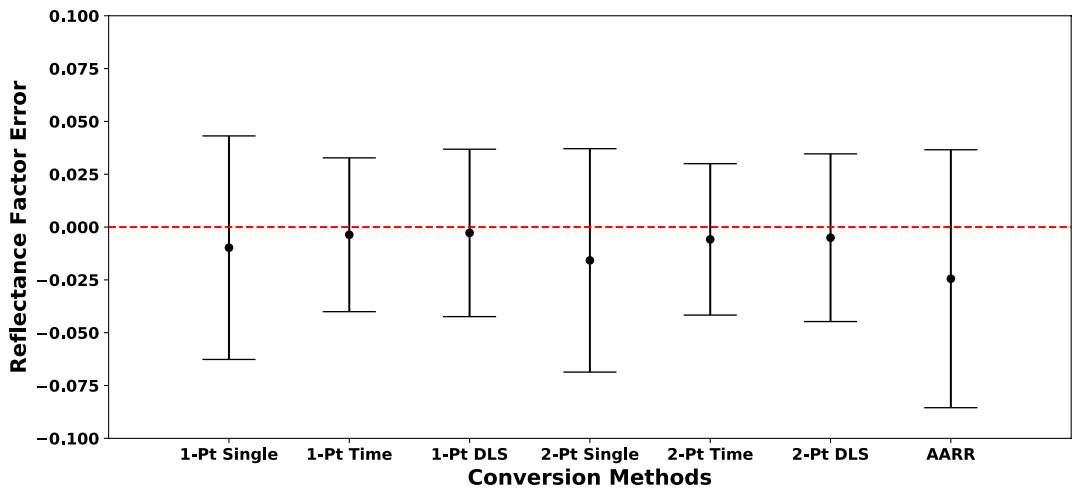

**Figure 13:** Overall average signed band effective reflectance factor errors displayed as box plots.

After analyzing the entire data set (9000+ data points across 1800+ targets across 1000+ images), the same conclusion was reached as the initial study [26]. ELM performed the best in producing signed (and absolute) reflectance factor errors across a variety of variables, while AARR performed very reliably.

### 5.4. M-AARR

Figures 14 and 15 show the downwelling radiance, sensor reaching radiance and reflectance curves for two of the 1920 simulations described in Table 2.

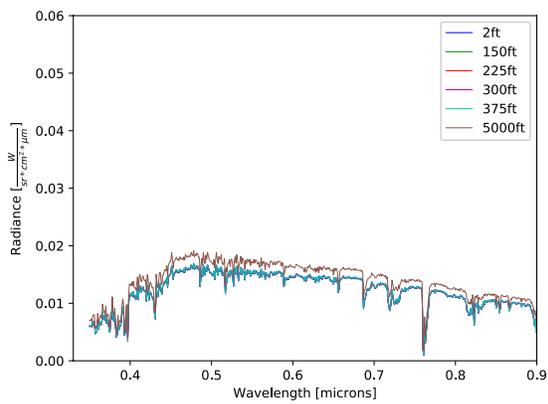
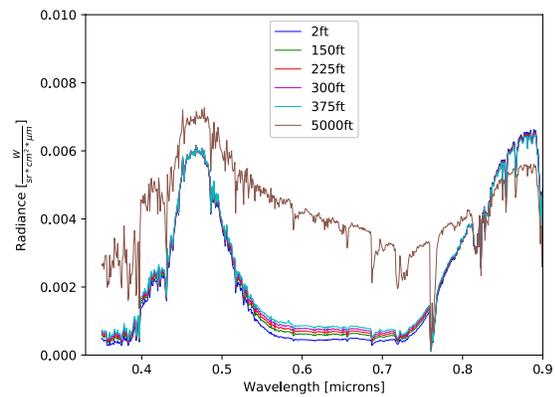
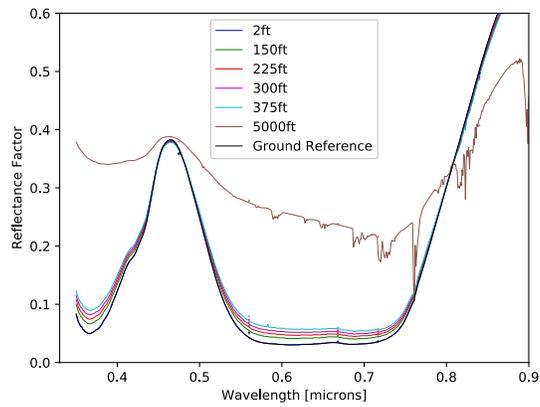

**Figure 14:** (a) Downwelling radiance (b) sensor reaching radiance and (c) reflectances simulated by MODTRAN for a blue felt target, at UTC, 5km visibility, day 79 and US standard atmosphere.

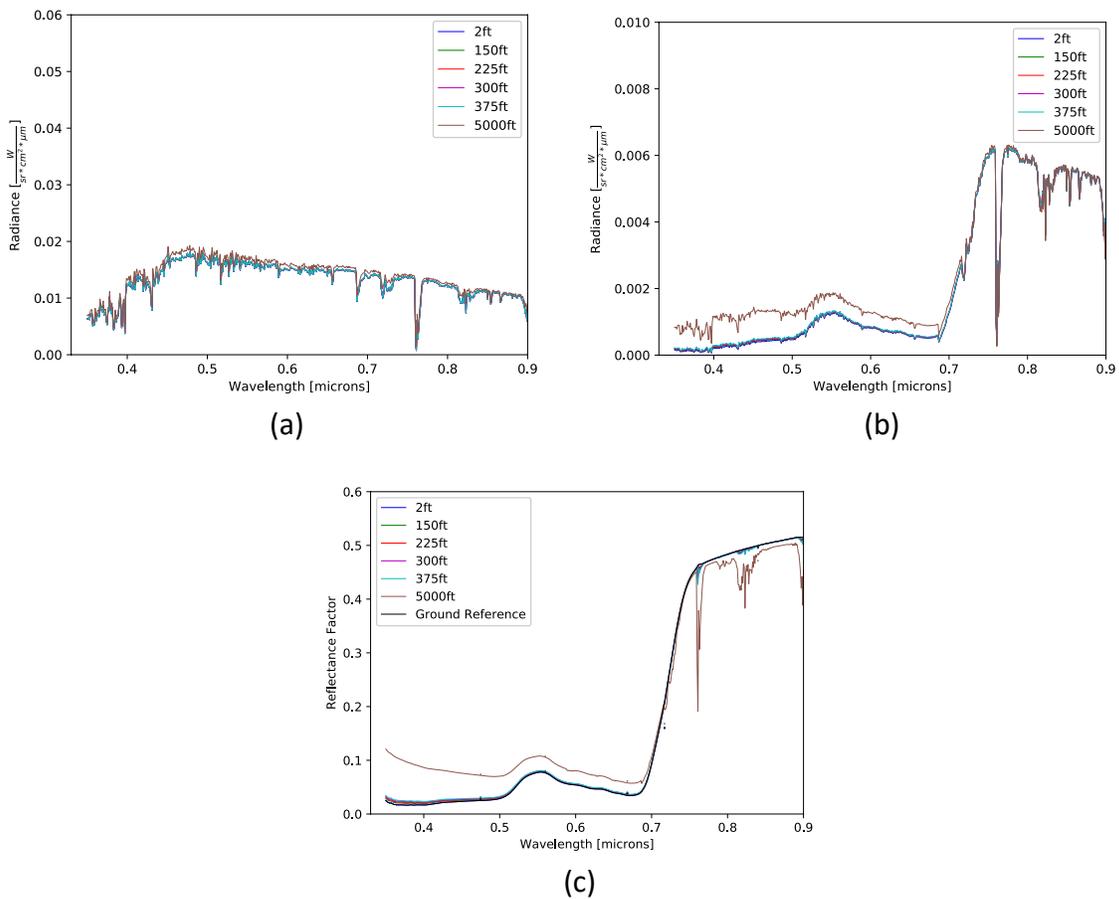

**Figure 15:** (a) Downwelling radiance (b) sensor reaching radiance and (c) reflectances simulated by MODTRAN for a grass target, at 18:00 UTC, 23km visibility, day 355 and mid-latitude winter atmosphere.

The overall average signed error in reflectance factor produced by M-AARR was 0.0023 with a standard deviation of 0.0074. To ensure the results from M-AARR were strictly depicting errors from typical sUAS altitudes, the 2 and 5,000ft simulations were excluded from these statistics. 2 and 5000ft were only used for figures to demonstrate how accurately the AARR technique can reproduce the reflectance curves (2ft), as well as demonstrate how poorly this technique works at traditional remote sensing heights (5,000ft).

Plots illustrating potential trends in average signed reflectance factor errors for these M-AARR simulated scenarios were generated across all the MicaSense RedEdge-3 sensor bands and variables. In Figure 16, time and atmosphere showed no significant variation in reflectance factor errors. Day number (season) did seem to have an effect on the reflectance factor. Figure 17 had the expected trends for visibility and sensor altitude. As the visibility increased, or the sUAS altitude decreased, both the errors and standard deviations decreased. This makes sense, as higher atmospheric visibility typically indicates a decrease in scattering particles in the atmosphere that can affect the path attenuation from the target to the sensor. In a similar fashion, as the sensor altitude decreases, the potential for attenuation and scattering decrease with decreasing path length, again resulting in more accurate prediction.

The plot for various targets in Figure 17(c) was more difficult to interpret. Each of the targets had different average errors and variability for their results. The smaller reflectance factor errors and standard deviations in reflectance factor errors for asphalt, concrete, and green felt are most likely a result of the overall small range in the average ground reference reflectance that were measured for those three targets. This can be seen in Table 3 or as illustrated in Figures 6-8. All of the targets start below 0.2 reflectance factor (at 0.4 microns), and while the reflectance factor of all the targets rise as the wavelength increases, there is no large inflection in reflectance factor for asphalt, concrete and green felt. Those three targets have an average ground reference reflectance factor close to 0.2 (at 0.9 microns). Blue felt, red felt and grass all reach much higher reflectance factor levels (greater than 0.5).

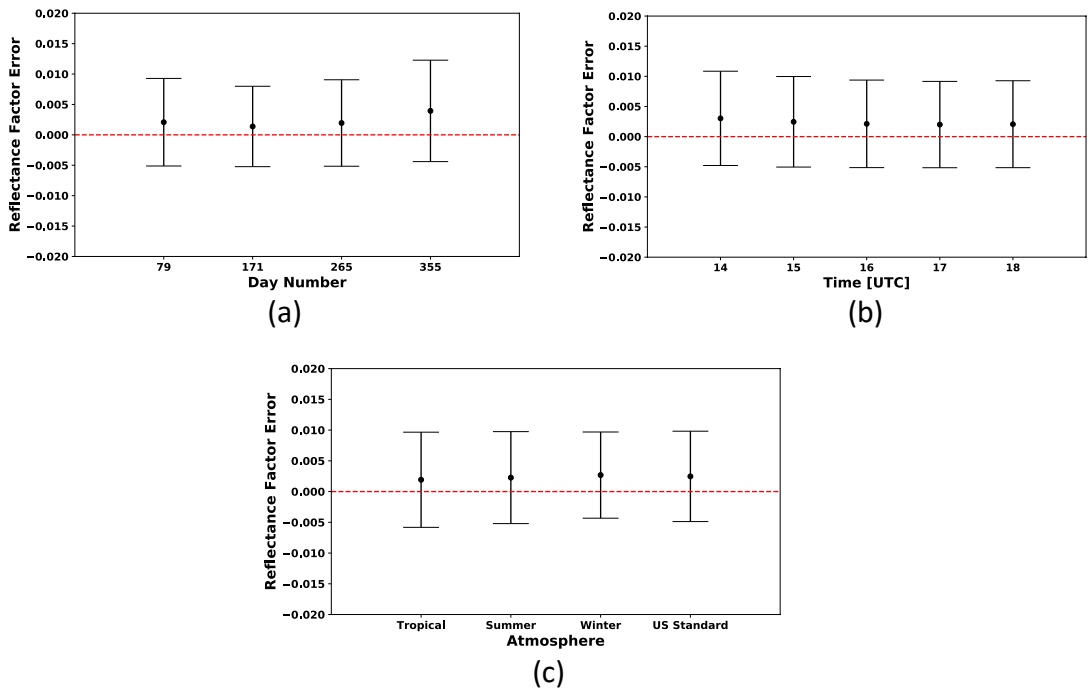

**Figure 16:** Signed reflectance factor errors averaged over MicaSense bands and variables, expect for (a) day number (b) time and (c) atmosphere.

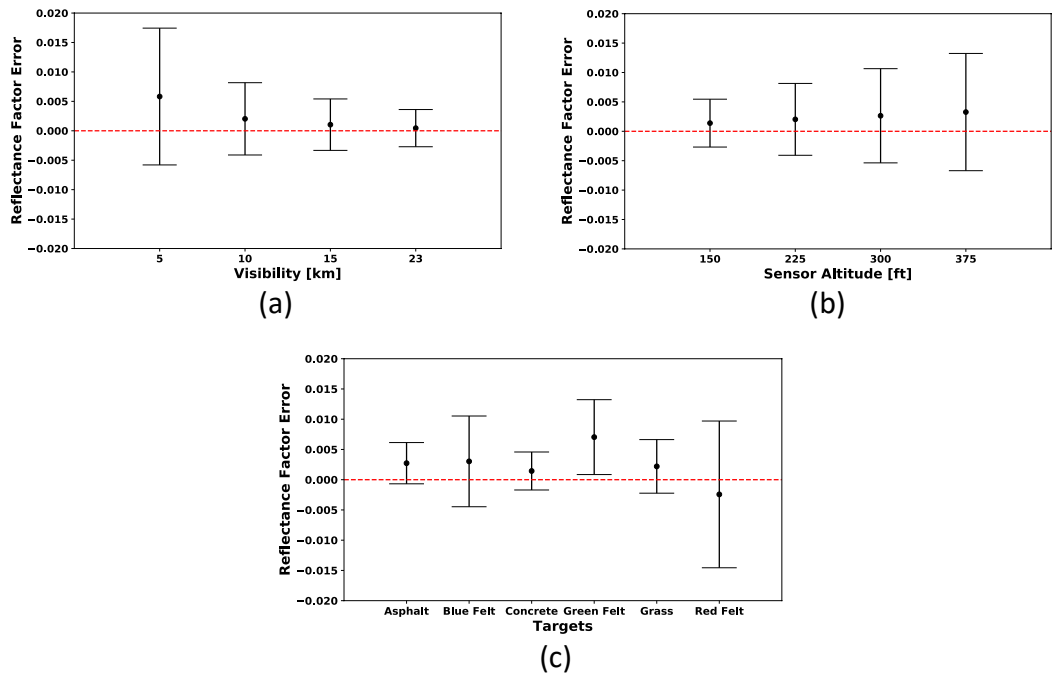

**Figure 17:** Signed reflectance factor errors averaged over MicaSense band and variables, except for (a) visibility (b) sensor altitude and (c) target

### 5.5. Overall

While the newly developed AARR technique for radiance to reflectance conversion does not surpass the results produced by ELM methods, it comes with a significant time save on sUAS field collections. Depending on the size of the collection field, and the number of reflectance conversion targets, setting up and measuring reflectance conversion targets can add significant time (45min - 1hour+) to field collects. If the field is large, reflectance conversion target revisits, which is important to produce the best results for both time and DLS ELM, can be burdensome as the sUAS might have to divert from its preplanned route. This will reduce the time that the sUAS is imaging the targets of interest, which decreases the area that can be covered by the sUAS. Every sUAS remote sensing practitioner has to determine which method is correct for their application. Research applications, specifically, require more accurate measurements to ensure the science behind the methods and technology is sound. ELM is the suggested approach for these types of work.

Because the results for all reflectance conversion methods were so close, a one-way ANOVA test of significance was carried out to determine if there was a significant difference. This produced a p value less than 0.001, which means the null hypothesis was not rejected, and there is a significant difference between the means of these conversion methods. However, the AARR method produces an averaged signed reflectance error of nearly -2.5%. If the analyzed target of interest has an inherent reflectance variability that is greater than 2.5%, then the error produced when converting radiance to reflectance would be overshadowed and ELM would arguably be excessive. This is because the best result that can be produced is directly correlated with the error present in the entire system (targets, camera, and processing). While ELM might never be fully replaced, AARR is a good stepping stone to sUAS image reflectance conversion without the use of panels.

## 6. Discussion

This in-depth study has shown that 2-point ELM produces the most accurate reflectance images, followed by 1-point ELM and AARR. The average reflectance factor errors were -0.0050 ± 0.0397, -0.0028 ± 0.0396, and -0.0244 ± 0.0610, respectively. These are all remarkably small errors for remote sensing imagery. While it did not outperform ELM, AARR was proven to be a viable method for converting sUAS remotely sensed images to reflectance, as the average reflectance factor error was slightly higher than ELM. These errors were computed across 1,820 targets, which implies strong validity to the findings. In addition, the expanded simulations of M-AARR also implies that AARR is a viable technique in a variety of atmospheres, days, times, heights, targets, and visibilities and that theoretically better results than those shown experimentally here might be achieved.

As with all experiments, errors in data collection impact the final results. Below is a discussion regarding the errors throughout the study.

### 6.1. Digital Count to Radiance Errors

After collecting the data, the raw imagery was converted to radiance before going to reflectance. As stated before, MicaSense's open source code was used for this step, however, the metadata values captured with every image (especially the coefficients) might only be applicable when the MicaSense RedEdge-3 is factory fresh. Constant use and handling of the camera impacts these coefficients, as does time since last calibration. Recent radiometric and geometric calibration could have reduced these errors further, and will be examined in upcoming studies.

### 6.2. Radiance to Reflectance Errors

Multiple parts of the radiance to reflectance conversion process may have introduced uncertainty. 1-point and 2-point ELM required ground reference reflectance curves, which were collected using a spectroradiometer. The collection of these reflectance curves may not have been identical every time (i.e. leveling of the spectrometer, variance in the position of the operator between standard and target radiance measurements, etc). In addition, the DLS provided by MicaSense might not have been an accurate measure of the current illumination conditions. A more precise sensor on top of the sUAS would have provided a better irradiance measurement which would have led to the computation of a more accurate reflectance image by the AARR technique.

## 7. Future Work
### 7.1. Digital Count to Radiance Conversion

If possible, prior to any data collection, a scientific imaging sensor should be radiometrically calibrated. This ensures that the user knows the variability between the data being collected and the data collected in a prior collection. A method for regular radiometric calibration is in development so errors in absolute radiance measurements may be more accurately determined. While this is not strictly required for reflectance factor generation, these more accurate radiance measurement will be available if determined necessary for any post-collection analyses.

### 7.2. Two Sensors Flying Together

One of the issues with our current data set was a lack of reflectance conversion images (for 1-point and 2-point ELM). It has been considered to fly two separate sUAS, each with a MicaSense RedEdge-3 attached. The first sUAS will fly the field as in this experiment, while the second sUAS hovers over a set of reflectance conversion panels for the entire duration of the mission. This would provide an image of the reflectance conversion panels at the same timing and interval as the field images. The only issue with this experiment is the cameras are not identical. Therefore, a model will need to be derived to cross calibrate the two MicaSense RedEdge-3 cameras or other spectral sensors.

### 7.3. Error Analysis

A complete error analysis will be conducted to analyze this entire process. This will help sUAS remote sensing practitioners known which part of processing produces the most error and how they can mitigate those errors. This error will include all measurements made by both the camera and the DLS, a particular target in the scene, the HR-1024i spectroradiometer, and any other field equipment used.

## 8. Conclusion

ELM proved to be the most accurate radiance to reflectance conversion methodology by producing an average signed reflectance factor error of -0.0050 ± 0.0397 (2-Point ELM) while AARR performed reliably (-0.0244 ± 0.0610). Depending on the accuracy required by the sUAS remote sensing practitioner, AARR can be used to produce reflectance imagery and save significant deployment and post-processing time. There would be no need to deploy large reflectance conversion panels for every mission, revisit these reflectance conversion panels multiple times, fly a second camera/downwelling sensor as suggested in the Future Work section, collect spectroradiometer data of these reflectance conversion panels, locate panels or develop complex code to process the data into reflectance.

**Acknowledgement**

The authors would like to thank Timothy Bauch, and Nina Raqueño for assisting with data collection (sUAS and spectroradiometer) in the field. The authors would also like to thank Geoffrey Sasaki, Ryan Connal, Kevin Kha, Jackson Knappen, Ryan Hartzell, and Evan Marcellus for assisting with manually locating ground targets in the sUAS images.

# References

[1] A. C. Watts, "Unmanned aircraft systems in remote sensing and scientific research: Classification and considerations of use," *Remote Sensing,* vol. 4, no. 6, pp. 1671-1692, 2012.

[2] C. H. Hegenholtz, "Small unmanned aircraft systems for remote sensing and earth science research," *Eos, Transactions American Geophysical Union,* vol. 93, 2012.

[3] I. Colomina, "Unmanned aerial systems for photogrammetry and remote sensing: A review," *ISPRS Journal of Photogrammetry and Remote Sensing,* vol. 92, pp. 79-97, 2014.

[4] C. Zhang, "The application of small unmanned aerial systems for precision agriculture: a review," *Precision Agriculture,* vol. 13, no. 6, pp. 693-712, 2012.

[5] D. J. Mulla, "Twenty five years of remote sensing in precision agriculture: Key advances and remaining knowledge gaps," *Biosystems Engineering,* vol. 114, no. 4, pp. 358-371, 2013.

[6] Agisoft, "Agisoft," Agisoft, [Online]. Available: http://www.agisoft.com/. [Accessed 4 8 2018].

[7] Agisoft, *Agisoft PhotoScan User Manual,* Agisoft, 2018.

[8] M. Ghazal, "UAV-based remote sensing for vegetation cover estimation using NDVI imagery and level sets method," in *2015 IEEE International Symposium on Signal Processing and Information Technology (ISSPIT)*, 2015.

[9] M. D. Biasio, "UAV-based Environmental Monitoring using Multi-spectral Imaging," *Proceedings of SPIE - The International Society for Optical Engineering,* 2010.

[10] B. Gao, "A Review of Atmospheric Correction Techniques for Hyperspectral Remote Sensing of Land Surfaces and Ocean Color," in *2006 IEEE International Symposium on Geoscience and Remote Sensing*, 2006.

[11] B. Gao, "Derivation of scaled surface reflectances from AVIRIS data," *Remote Sensing of Environment,* vol. 44, pp. 165-178, 1993.

[12] F. Kruse, *Comparison of ATREM, ACORN, and FLAASH atmospheric corrections using low-altitude AVIRIS data of Boulder, CO,* 2004.

[13] Q. Zheng, "The High Accuracy Atmospheric Correction for Hyperspectral Data (HATCH) model," *IEEE Transactions on Geoscience and Remote Sensing,* vol. 41, pp. 1223-1231, 2003.

[14] S. Zhu, "Retrieval of Hyperspectral Surface Reflectance Based on Machine Learning," *Remote Sensing,* vol. 10, no. 2, p. 323, 2018.

[15] S. Diek, "Creating Multi-Temporal Composites of Airborne Imaging Spectroscopy Data in Support of Digital Soil Mapping," *Remote Sensing,* vol. 8, 2016.

[16] G. M. Smith, "The use of the empirical line method to calibrte remotely sensed data to reflectance," *International Journal of Remote Sensing,* vol. 20, no. 13, pp. 2653-2662, 1999.

[17] C. Wang, "A Simplified Empirical Line Method of Radiometric Calibration for Small Unmanned Aircraft Systems-Based Remote Sensing," *IEEE Journal of Selected Topics in Applied Earth Observations and Remote Sensing,* vol. 8, no. 5, pp. 1876-1885, 2015.

[18] F. Kruse, "Mineral mapping at Cuprite, Nevada with a 63-channel imaging spectrometer," *Photogrammetric Engineering and Remote Sensing,* vol. 56, no. 1, pp. 83-92, 1990.

[19] J. L. Dwyer, "Effects of empirical versus model-based reflectance calibration on automated analysis of imaging spectrometer data: a case study from the Drum Mountains, Utah," *Photogrammetric Engineering and Remote Sensing,* vol. 61, no. 10, pp. 1247-1254, 1995.

[20] A. S. Laliberte, "Multispectral remote sensing from unmanned aircraft: Image processing workflows and applications for rangeland environments," *Remote Sensing,* vol. 3, no. 11, pp. 2529-2551, 2011.

[21] E. Bondi, "Calibration of UAS imagery inside and outside of shadows for improved vegetation index computation," *International Society for Optics and Photonics,* vol. 9866, 2016.

[22] W. H. Farrand, "Retrieval of apparent surface reflectance from AVIRIS data: A comparison of empirical line, radiative transfer, and spectral mixture methods," *Remote Sensing of Environment,* vol. 47, no. 3, pp. 311-321, 1994.

[23] R. Price, "Preliminary evaluation of CASI preprocessing techniques," in *Proceedings of the 17th Canadian Symposium on Remote Sensing*, 1995.

[24] M. S. Moran, "A refined empirical line approach for reflectance factor retrieval from Landsat-5 TM and Landsat-7 ETM+," *Remote Sensing of Environment,* vol. 78, pp. 71-82, 2001.

[25] E. Karpouzli, "The empirical line method for the atmospheric correction of IKONOS imagery," *International Journal of Remote Sensing,* vol. 24, no. 5, pp. 1143-1150, 2003.

[26] B. Mamghani, "An initial exploration of vicarious and in-scene calibration techniques for small unmanned aircraft systems," *arXiv preprint arXiv:1804.09585,* 2018.

[27] T. Hakala, "Direct Reflectance Measurements from Drones: Sensor Absolute Radiometric Calibration and System Tests for Forest Reflectance Characterization," *Sensors,* 2018.

[28] J. Lekki, "Development of Hyperspectral remote sensing capability for the early detection and monitoring of Harmful Algal Blooms (HABs) in the Great Lakes," in *AIAA Infotech Aerospace Conference and AIAA Unmanned Unlimited Conference*, 2013.

[29] J. D. Ortiz, "Intercomparison of Approaches to the Empirical Line Method for Vicarious Hyperspectral Reflectance Calibration," *Frontiers in Marine Science,* vol. 4, p. 296, 2017.

[30] MicaSense, *MicaSense Manual,* MicaSense Incorporated, 2015.

[31] S. Cao, "Radiometric calibration assessments for UAS-borne multispectral cameras: Laboratory and field protocols," *ISPRS Journal of Photogrammetry and Remote Sensing,* vol. 149, 2019.

[32] J. Schott, Remote Sensing: The Image Chain Approach, Oxford University Press, 2007.

[33] M. Incorporated, *Image Processing,* https://github.com/micasense/imageprocessing, 2017.

[34] SSI, "MODTRAN," SSI, [Online]. Available: http://modtran.spectral.com/modtran_index. [Accessed 27 3 2017].

[35] Berk, *MODTRAN4 User's Manual,* Air Force Research Lab, 1999.

[36] G. 8. Tech, "Group8tech," [Online]. Available: https://www.group8tech.com/. [Accessed 24 07 2018].

[37] FAA, "FAA," [Online]. Available: https://www.faa.gov/uas/. [Accessed 06 10 2018].

[38] S. V. Corporation, *Spectral Vista Corporation, HR-1024i,* https://www.spectravista.com/hr-1024i/, 2017.

[39] C. J. Bruegge, "Use of Spectralon as a diffuse reflectance standard for in-flight calibration of earth-orbiting sensors," *Optical Engineering,* vol. 32, 1993.

[40] Labsphere, *Spectralon Diffuse Reflectance Targets,* Labsphere, 2017.

[41] D. McDowell, "Spectral Distribution of Skylight Energy for Two Haze Conditions," *Photogrammetric Eng,* vol. 40, pp. 569-571, 1974.